\documentclass[aps,groupedaddress,preprint]{revtex4-1}
\usepackage{graphicx}
\usepackage{xcolor}
\usepackage{caption}
\usepackage[utf8]{inputenc}
\usepackage[percent]{overpic}
\begin{document}


\title{Near-IR Photoluminescence of C$_{60}^+$ and Implications for Astronomy}


\author{Dmitry Strelnikov}
\email[]{dmitry.strelnikov@kit.edu}
\author{Bastian Kern}
\altaffiliation[current address: ]{Max Planck Institute for Solid State Research, Heisenbergstraße 1, 70569 Stuttgart, Germany}
\author{Manfred M. Kappes}
\affiliation{Karlsruhe Institute of Technology (KIT), Division of Physical Chemistry of Microscopic Systems, Karlsruhe, Germany}


\date{\today}
 
\begin{abstract}
We have observed that C$_{60}^+$ ions isolated in cryogenic matrixes show pronounced near-IR photoluminescence upon excitation in the near-IR range. By contrast UV photoexcitation does not lead to measurable luminescence. After the recent unequivocal assignment of five Diffuse Interstellar Bands to near-IR absorption bands of C$_{60}^+$, we propose to search also for C$_{60}^+$ near-IR emission in those astronomical objects, where fullerenes have been already detected or may be potentially present.
\end{abstract}


\maketitle

\begin{figure*}[h]
\includegraphics[width=0.9\textwidth]{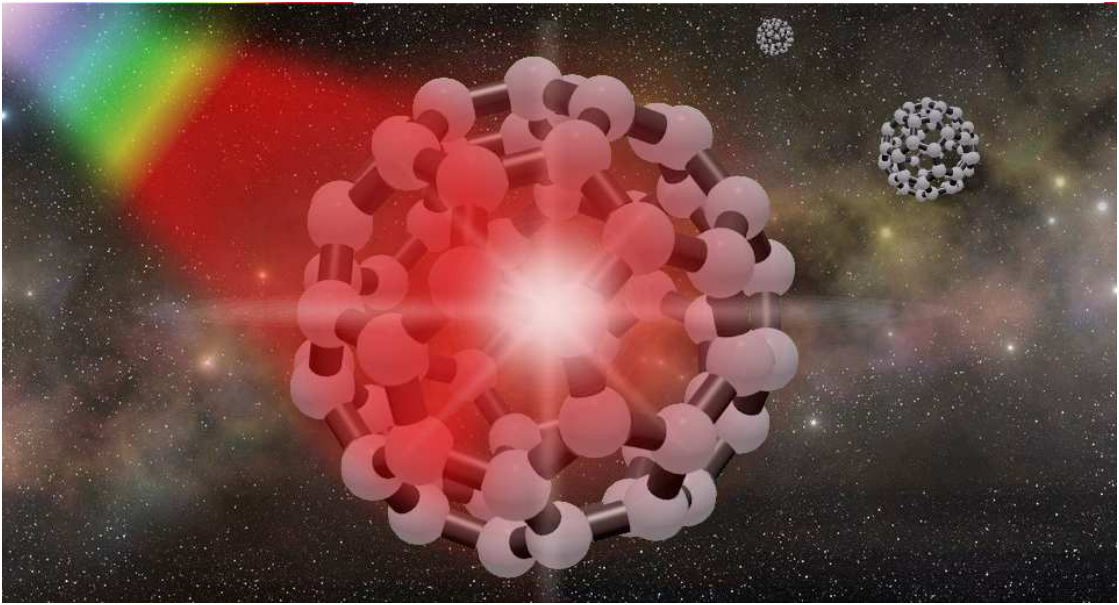}
\caption*{TOC Figure: C$_{60}^+$ strongly emits Stokes-shifted near-IR photons after near-IR excitation. The schematic drawing indicates a probable scenario in which the UV component of the excitation light is scattered away by dust particles.}
\end{figure*}
\section*{}

\textbf{Recently, there has been renewed interest in fullerenes   in several research fields including astronomy, astrophysics, physical chemistry and quantum chemistry. This was triggered by the detection of the neutral forms of C$_{60}$ and C$_{70}$ in circumstellar environments\cite{C60_space1,C60_space2,C60_space3,C60_space2013} and of charged C$_{60}^+$ in the diffuse interstellar medium\cite{MaierDIBs2015,MaierDIBs2015_2}. The firm identification of C$_{60}$ and C$_{70}$ was based on their vibrational infrared emissions, while C$_{60}^+$ was identified by its electronic absorptions in the near-IR region. As a corollary, C$_{60}^+$ has turned out to be responsible for several of the known but previously unassigned diffuse interstellar bands (DIBs)\cite{DIBs}. Thus a first significant step towords solving an almost 100 years old puzzle in the spectroscopical and astronomical communities has now been taken. There are also hints for the presence of C$_{60}^{+/2+}$ in the 
circumstellar environments, based on the infrared spectral data\cite{JoblinC60+,Depo2C60AA}.} 

\textbf{Most of the UV-vis and near-IR region spectral bands observed in circumstellar environments are absorption features. The light source in this case is simply the  star behind the clouds, containing the absorbing molecules. On the other hand, most of the corresponding astronomical spectral features observed in the mid-IR region are emission bands. Molecular-like emission in the shorter wavelength range of electronic transitions is not a frequently observed phenomenon in astronomy. There are exceptions: one of the most prominent objects, having emission lines in the visible range is the 'Red Rectangle' nebula in the Monoceros constellation\cite{RedRectangle2002}.} 

\textbf{In our laboratory investigation of fullerene ions, we found that C$_{60}^+$ manifests pronounced  fluorescence in the near-IR, when also photo-excited in the near-IR. We suggest therefore to look for the characteristic near-IR emission features of C$_{60}^+$  in astronomical environments, where either  C$_{60}$/C$_{60}^+$  has already been detected or where fullerenes are likely to  be present. Below we provide the first quantitative experimental data characterizing this emission.}

\section*{}
\begin{figure*}
\includegraphics[width=0.7\textwidth]{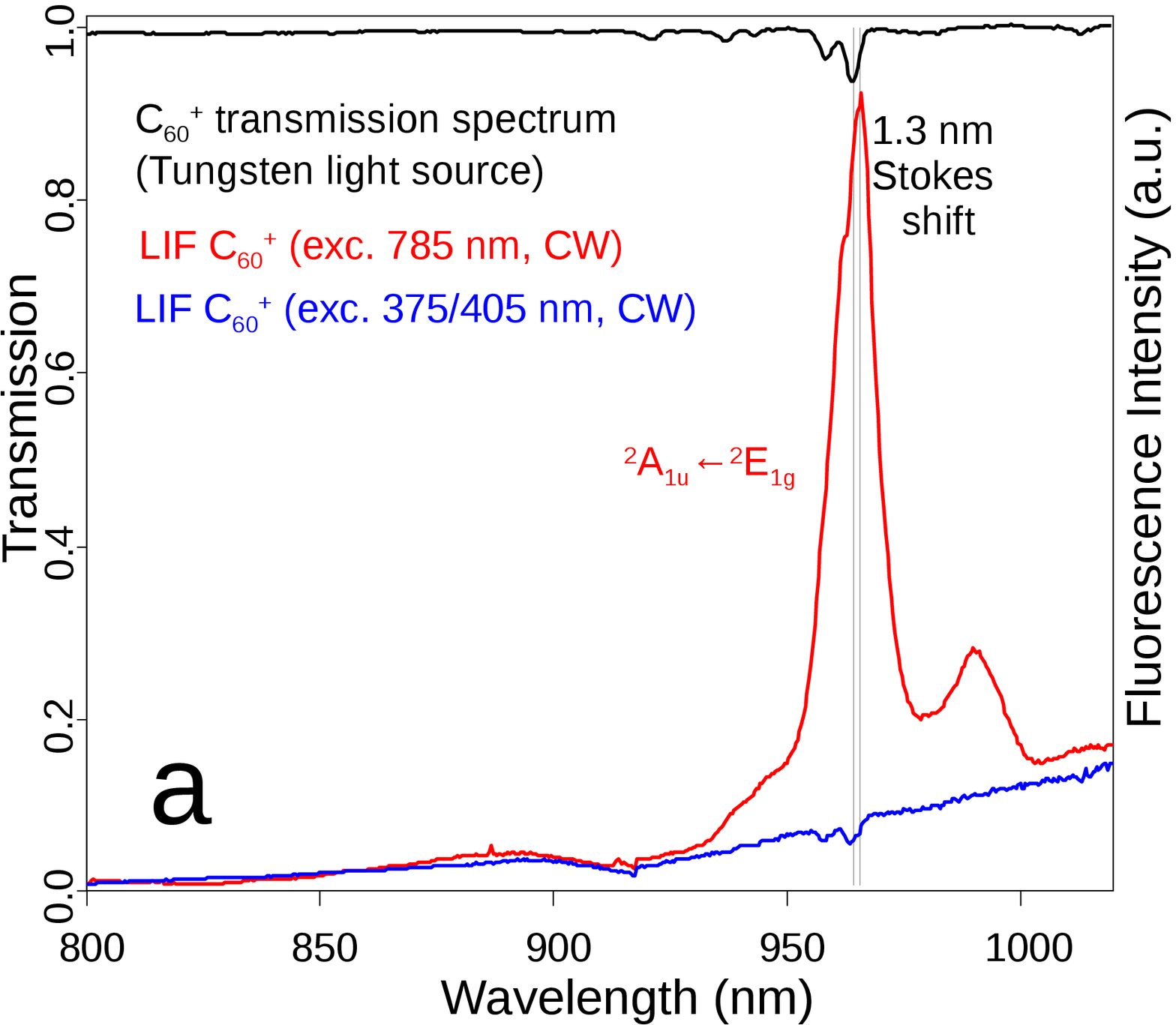}%
\hspace{0.01\textwidth}
\raisebox{0.3\height}{\includegraphics[width=0.25\textwidth]{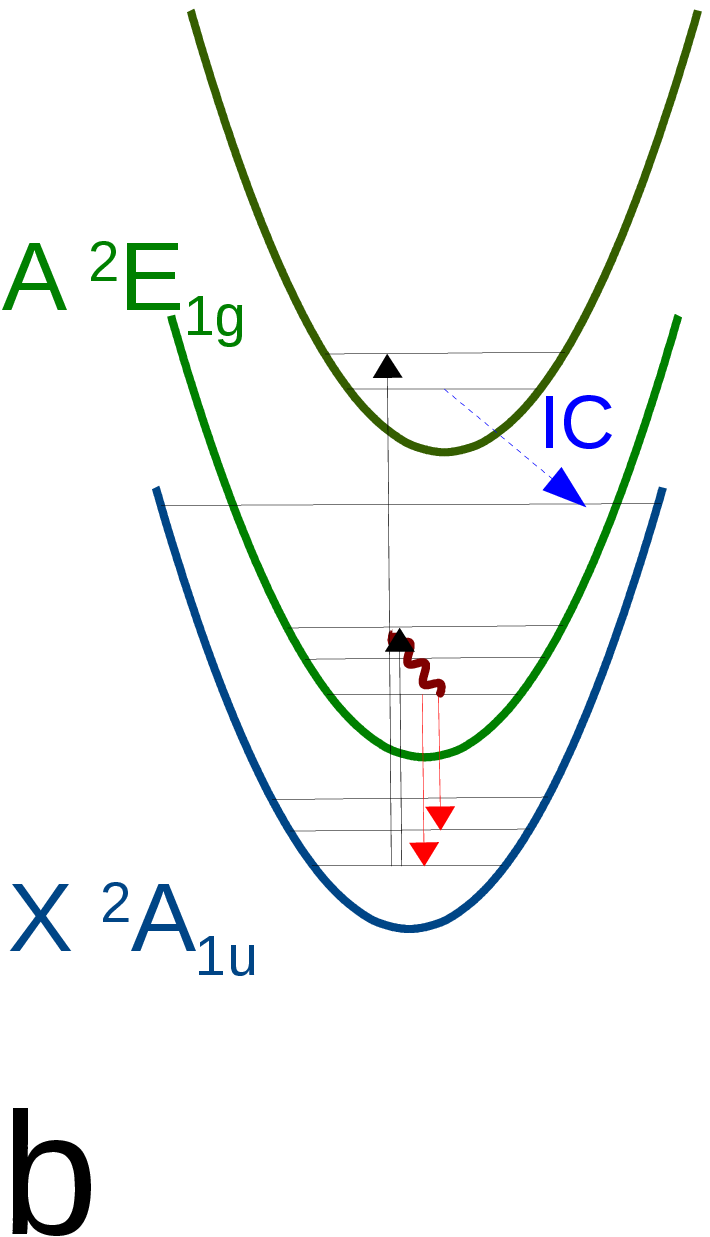}}%
\caption{\label{LIF1D} \textbf{(a)} Spectroscopic measurements on C$_{60}^+$ containing cryogenic neon matrixes as indicated. \textbf{(b)} Energy level diagram rationalizing these observations in terms of radiative (red arrows) and non-radiative decays.}
\end{figure*}
Fig.~\ref{LIF1D}a shows three spectroscopic measurements  carried out on the same sample, containing known concentrations of C$_{60}^+$, C$_{60}^-$ and C$_{60}$ isolated in a Ne matrix at 5~K: (i) a transmission spectrum , (ii) a laser-induced dispersed fluorescence (LIF) spectrum obtained  upon excitation at 785~nm and (iii) a corresponding LIF measurement upon laser excitation at 405nm. LIF measurement (ii) shows an emission band  at 965~nm which is Stokes shifted by about 1.3~nm from the 0-0 transition of C$_{60}^+$ in neon. This band can be observed only in samples, containing C$_{60}^+$. One can also recognize a LIF transition to the first vibrationally excited level of the electronic ground state shifted by about 25~nm  (262~cm$^{-1}$) to lower energies. This  fits to a  radial vibrational mode of C$_{60}^+$  -- probably, the quadrupolar deformation mode A$_{1g}$ calculated to lie at 258~cm$^{-1}$ for the isolated molecule (harmonic frequency, RI-DFT BP86/def-SV(P) level of theory)\cite{Turbomole}. Laser excitation at shorter  wavelengths in the UV region (405~nm and also 375~nm (not shown)) does not lead to detectable  emission at 965~nm. Instead we  observe a weak featureless emission slowly increasing into the IR. The carrier of the latter is unclear. It is however only observed for fullerene containing matrixes and may reflect multiphoton absorption processes. While matrix isolated C$_{60}$ is known to show broad weak photoluminescence upon UV excitation\cite{C60lumin0,C60lumin,C60lumin1}, there have been no previous reports of light emission from C$_{60}^+$. 

\begin{figure*}
\includegraphics[width=0.64\textwidth]{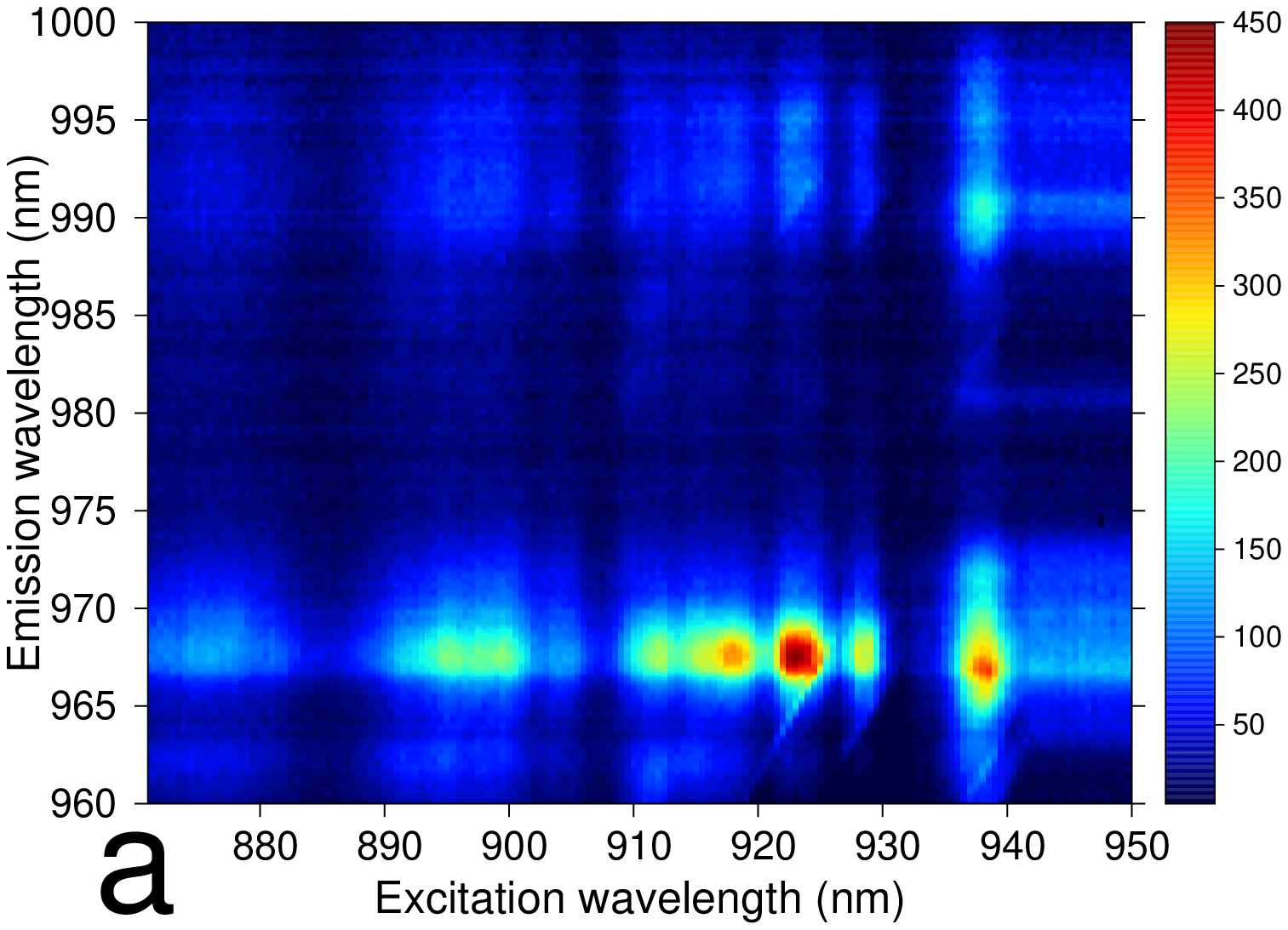}%
\raisebox{0.05\height}{\includegraphics[width=0.34\textwidth]{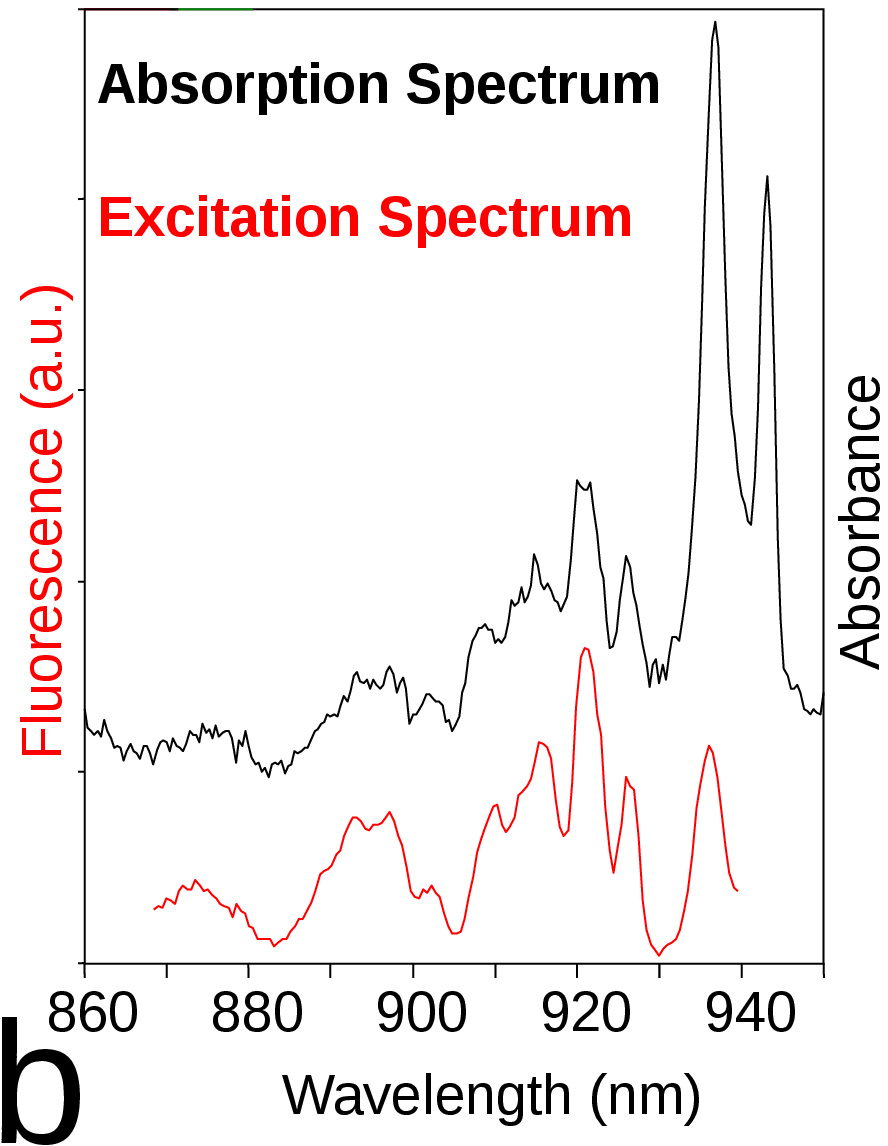}}%
\caption{\label{LIF2D} \textbf{(a)} Fluorescence excitation/emission map obtained by tunable pulsed laser irradiation of a C$_{60}^+$ containing neon matrix. \textbf{(b)} Fluorescence excitation spectrum (integrated emission as a function of excitation wavelength) in comparison to the absorption spectrum of C$_{60}^+$.}
\end{figure*}
To confirm that the 965~nm emission really originates from C$_{60}^+$, we also used tunable pulsed laser excitation from 800~nm to 940~nm in order to record a photoluminescence excitation spectrum -- this wavelength region contains the known electronic absorptions $^2$E$_{1g}\leftarrow$X$^2$A$_{1u}$ of C$_{60}^+$ \cite{C60ions_matrix_JPMaier,C60cC70cMaierGasPhase,Depo2C60AA}. The corresponding photoluminescence map is displayed in Fig.~\ref{LIF2D}a. A comparison of the photoluminescence excitation and absorption spectrum is shown in Fig.~\ref{LIF2D}b. For the excitation spectrum the integrated intensity of the 965~nm band was used as an ordinate. It is apparent, that the spectra are very similar and therefore that the 965~nm emission band is produced by C$_{60}^+$ upon population of its first exited state. 

We also checked the laser power dependence of the 965~nm emission band intensity. This measurement revealed  a one photon process.     We have also determined an  upper limit for the lifetime of the excited state to be 2~ns. The fluorescence quantum yield for C$_{60}^+$ absorptions in the range [775;~971]~nm was estimated to be $\Phi\gtrsim 4.2\cdot 10^{-4}$. All these measurements are detailed in Supporting Information.

From the astronomical point of view it is important to know the gas-phase position of this emission. The 0-0 electronic absorption transition of C$_{60}^+$ has been found to lie at  963.2~nm in gas phase\cite{MaierDIBs2015}. Assuming, that the Stokes shift is similar to that in the Ne-matrix environment, we predict the position of the gas-phase emission maximum to be 964.5~nm. Regarding the line width, it is most likely similar to the absorption line width which was found to be 2--3~\AA~for isolated C$_{60}^+$ held in an 4~K ion trap\cite{MaierDIBs2015}. Neutral diffuse clouds and molecular clouds potentially containing C$_{60}^+$ can have temperatures in the range 10--100~K. Therefore we also explored the temperature dependence of C$_{60}^+$ fluorescence by performing analogues measurements in Kr matrix. These measurements at various temperatures up to 50~K revealed no significant changes of LIF intensity (below 7\%). By extrapolation we expect  that C$_{60}^+$ still measurably photoluminesces in the near-IR at temperatures up to 100~K.

Taking into account all experimental information (position, linewidth, NIR single-photon excitation, quantum efficiency) we propose to search for the C$_{60}^+$ emission in those astronomical objects, for which C$_{60}$/C$_{60}^+$ has already been detected. Particularly promising from this point of view would be those objects in which only NIR excitation can take place (simultaneous UV excitation would produce additional background from fullerenes and other species present, thus decreasing the contrast and making detection of the C$_{60}^+$ NIR emission signal more difficult). One of the candidate objects could be the reflection nebula \textit{CED~201}, where neutral C$_{60}$ has been already detected \cite{C60_CED201}. This object shows unidentified luminescence bands in the visible range (465--752~nm)\cite{Simonia2004}, which may possibly extend into the near-IR.
 \section*{Methods}
\subsection*{Sample preparation}
The UHV experimental setup for optical spectroscopy of mass-selected species has  already been described in our previous publications\cite{Depo2C60,Depo2C60dicat,Depo2C60AA}. Briefly, C$_{60}^+$ ions are produced in an electron-impact ionization source, guided through an electrostatic lens system, mass selected by a quadrupole mass filter and co-deposited with neon or other rare gases onto a 5~K Al- or Au-coated sapphire substrate. The ion current of mass-selected C$_{60}^+$ is usually 150-250~nA. Despite our only depositing  C$_{60}^+$, the matrix formed also contains  C$_{60}$ and C$_{60}^-$ which are generated by neutralization and charge changing processes during soft landing of the cations. Under typical deposition conditions the resulting matrixes  contain an approximate ratio of (20\%C$_{60}^+$)/(60\%C$_{60}$ )/(20\%C$_{60}^-$)\cite{Depo2C60}. For the measurements described we deposited about 10$^{14}$ C$_{60}^+$ ions.
\subsection*{LIF Spectroscopy}
\begin{figure}
\includegraphics[width=0.4\textwidth]{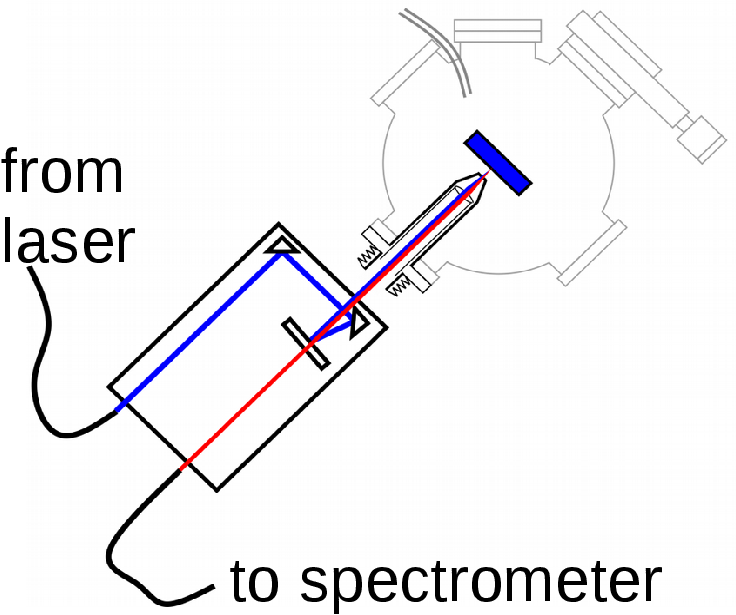}
\caption{\label{LIFsetup} Experimental configuration for fluorescence measurements of C$_{60}^+$ in neon.}
\end{figure}

For photoexcitation we used CW diode lasers at 375, 405, 785~nm as well as  a 10~ns pulse width OPO laser running at a repetition rate of  20~Hz (and tunable in the 870--950~nm range (idler beam of a Coherent Panther OPO laser)). The sample spot diameter was about 10~mm. The laser beam diameters at sample were about 1~mm for the 375, 405 and the OPO lasers and 100~$\mu$m for the 785~nm laser. Typical iradiation powers were: 20~mW for the 375~nm laser, 20--200~mW for the 405~nm laser, 10-250~mW for the 785~nm laser, and about 1~mW for the pulsed laser. A 950~nm shortpass filter was used to clean up the tunable pulsed laser excitation light.  A Raman spectrometer (Kaiser Optics RXN1) with a 950~nm longpass filter (Thorlabs FELH0950) was used to detect the emission. The corresponding excitation and collection geometry was perpendicular to the substrate, Fig.~\ref{LIFsetup}.

\subsection*{Power dependence}
\begin{figure}
\includegraphics[width=0.5\textwidth]{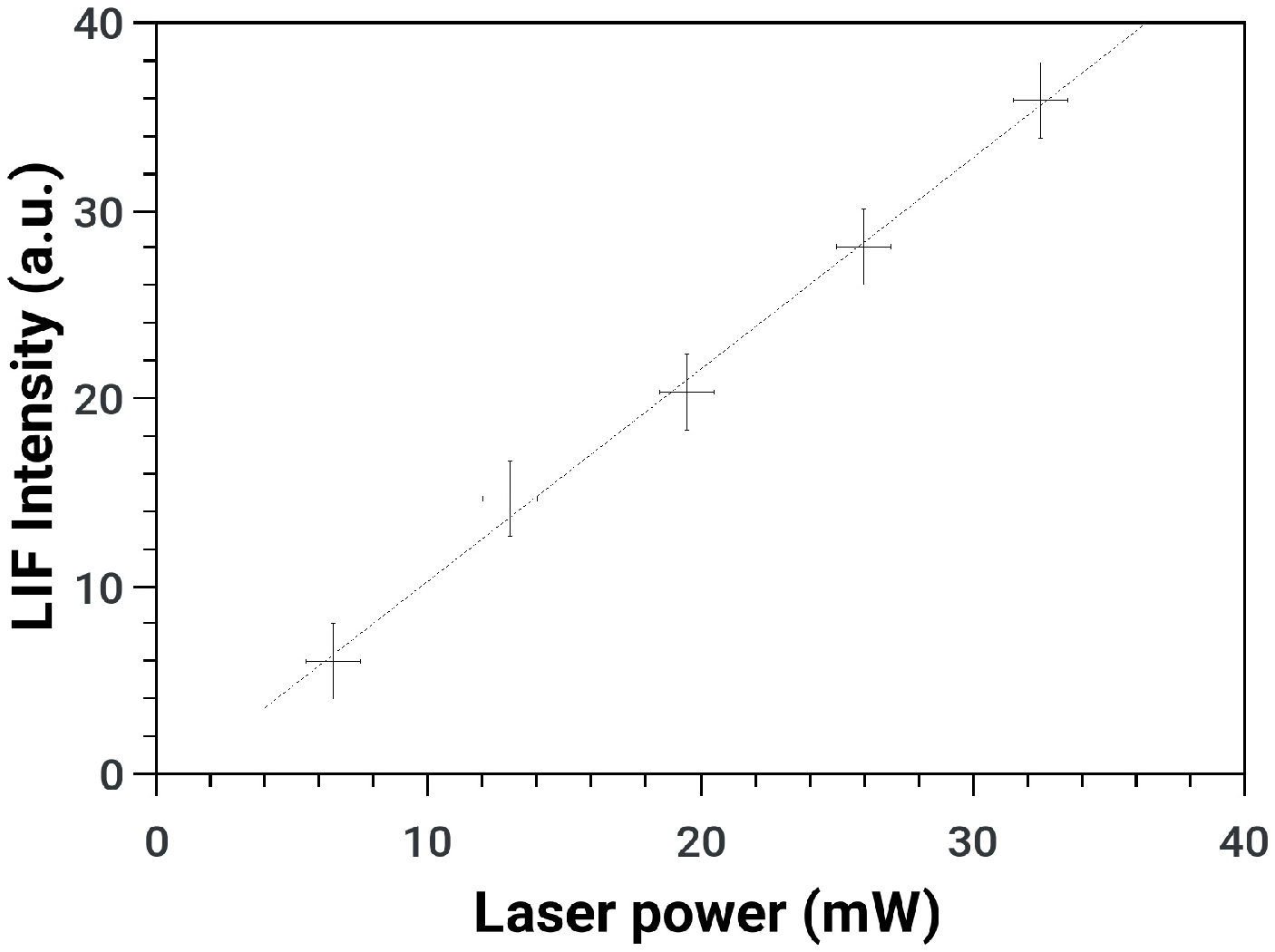}
\caption{\label{LIFpower} Total fluorescence intensity versus laser power (785~nm excitation).}
\end{figure}

Power dependence measurements were performed using the 785~nm CW diode laser. We found that the matrix samples start to degrade upon excitation with more than 150~mW of laser power. Below this power level LIF signal intensity remains  reproducible when switching back and forth between different excitation powers. For excitation powers below 40~mW, the LIF intensity clearly showed a linear dependence, Fig.~\ref{LIFpower}, indicating a one-photon process.

\subsection*{Quantum efficiency}
To estimate the quantum efficiency of the C$_{60}^+$ NIR emission, one has to know the number  of excitation photons, the C$_{60}^+$ quantity and absorption cross section  at the excitation wavelength -- as well as  the corresponding number  of emitted photons.

The accuracy to which some of these quantities can be determined is limited by instrumental factors which can be best taken into account by applying a reference standard. As a reference substance we took benzene. Its non-resonant Raman cross-section for an excitation at 785~nm can be extrapolated from experimental data\cite{BenzeneRaman}: $(\frac{d\sigma}{d\Omega})_{benzene}\approx 5\cdot 10^{-30}[cm^2Molec^{-1}Sr^{-1}]$ for the 992~cm$^{-1}$ Raman line. The Raman cross-section is defined as ${\frac{d\sigma}{d\Omega}=P_R/I_{ex}}$, where $P_R$ is the power of the Raman emission and  $I_{ex}[W\cdot cm^{-2}]$ is the power density of the excitation light. The specifications of the Raman MKII Probe Head with NCO-1.3-VIS/NIR optics are as follows: the collection zone diameter is $d=100~\mu$m, the collection zone depth is $l=500~\mu$m, and the light-collection solid angle of the objective is $\Omega\approx0.2~$Sr. We accumulated Raman emission of benzene for $\Delta t=1$~s, exciting 
with the $P_{ex}=1$~mW  CW laser at $\lambda_{ex}=785$~nm. The total number of emitted photons collected by the objective is then\[N_{photons}=\frac{d\sigma}{d\Omega}\Omega\frac{P_{ex}}{\pi (\frac{d}{2})^2}\Delta t \frac{\lambda_{ex}}{hc} N_{
molecules},\] where $N_{molecules}$ is the number of benzene molecules in the collection volume. For the listed parameters this corresponds to $N_{photons}\approx 1.5\cdot 10^6$ for the benzene 992~cm$^{-1}$ Raman line. This number of photons corresponds to  1730 intensity counts at 851~nm (992~cm$^{-1}$ Raman shift) as registered by  the CCD detector and displayed by the spectrometer software.

We deposited about 5000~nA$\cdot$min of C$_{60}^+$ to create a neon matrix, containing C$_{60}^{+/-/0}$. From our previous experiments, we know that the charge distribution is roughly 20:20:60 for +/-/0 respectively. Therefore, the amount of C$_{60}^+$ is about $3.75\cdot 10^{14}$ ions. These ions are almost homogeneously distributed over a spot about 10 mm in diameter (as experimentally determined). Correspondingly,  the number of C$_{60}^+$ ions in the Raman signal collection zone is (100~$\mu m$ diameter) about $3.75\cdot 10^{10}$. Exciting this amount of C$_{60}^+$ with a 1~mW laser and accumulating for 1~s gives rise to  $7.5\cdot 10^3$ (intensity corrected, because the reference substance emission is at 851~nm, where the CCD has larger sensitivity) intensity counts or $3.2\cdot 10^{6}$ photons of  C$_{60}^+$ emission. The total number of emitted photons in all directions would be about $2\cdot10^{8}$ corresponding to the 1~mW excitation/1~s accumulation parameters.


The number of the absorbed photons at wavelength $\lambda$  can be derived using the Lambert-Beer law and relations between oscillator strength $f_\lambda$ and absorption cross section at this wavelength\cite{Conversions}: \[ N_{absPh}=N_{exPh}\left(1-exp\left(-\frac{e^2}{\varepsilon_0 m_ec\Delta\omega_\lambda} f_{\lambda}\frac{N_{C60+}}{\pi {(d/2)}^2}\right)\right),\]
where $\Delta\omega_\lambda$ ($rad/s$) is the fwhm of the absorption at the wavelength $\lambda$. 

The oscillator strength of the C$_{60}^+$ absorption at 785~nm is hard to estimate, because it can not be clearly distinguished from the noise at this wavelength, therefore we set only an upper limit: $f_{785}\leq 2\cdot 10^{-5}$. In this wavelength range, stronger and therefore more clearly distinguishable C$_{60}^+$ absorptions in Ne matrix have roughly the same linewidths of about 50~cm$^{-1}$. We assume that the absorption at 785~nm also has such a linewidth. With these assumptions, the quantum yield for the excitation at 785~nm is $\Phi_{785}\geq 4.2\cdot 10^{-4}$. In a first approximation the quantum yield for other C$_{60}^+$ transitions in the $^2E_{1g}\leftarrow^2A_{1u}$ vibronic system is expected to be quite similar. 
Experimental data for other $f_\lambda$-numbers in neon and gas-phase cross-sections can be found elsewhere\cite{Depo2C60AA,C60cC70cMaierGasPhase}.

\subsection*{Lifetime estimation}
\begin{figure}
\includegraphics[width=0.5\textwidth]{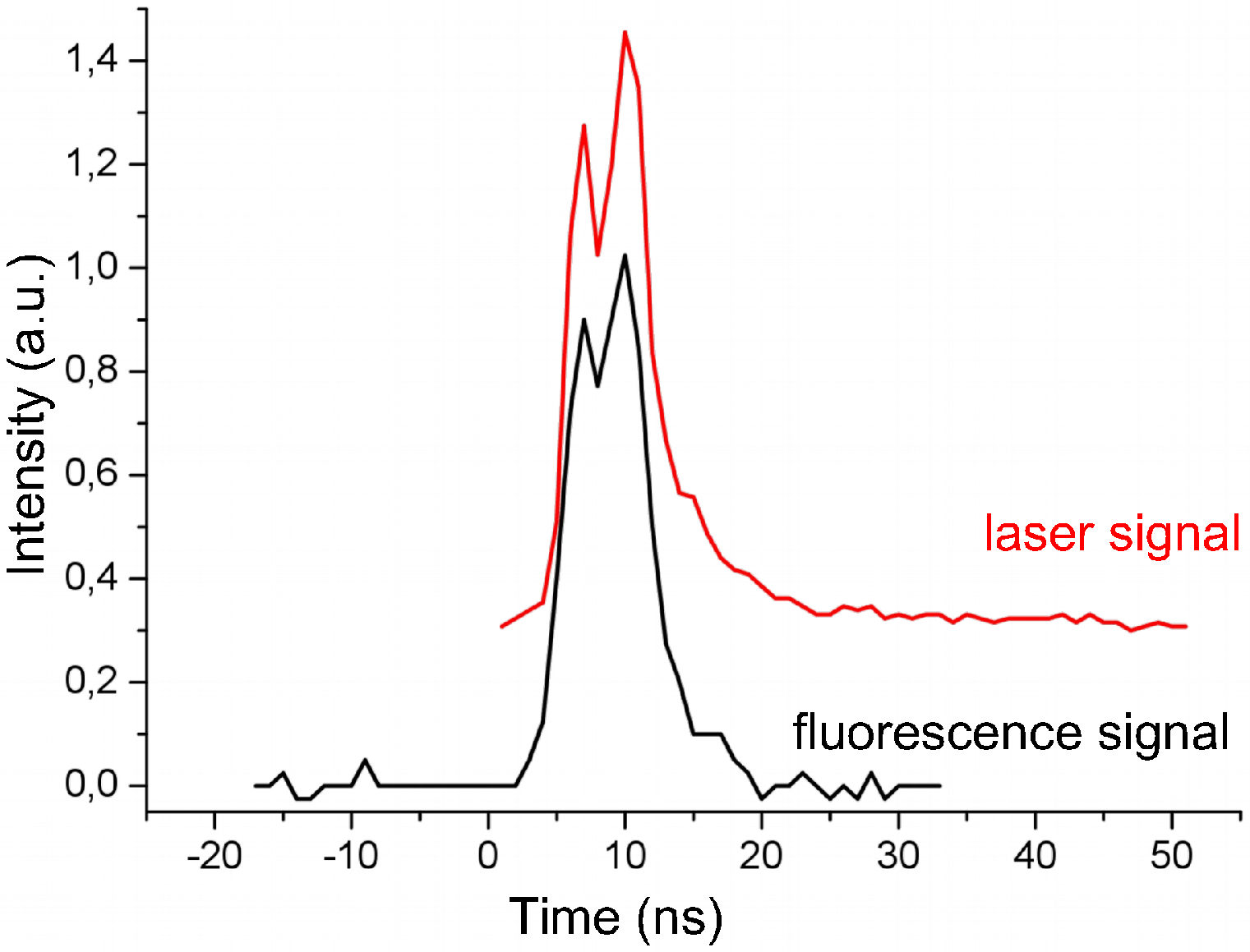}
\caption{\label{LIFPMT} Time dependence of the fluorescence signal versus that of the pulsed laser excitation source (923~nm). Note the strong similarity indicating that the fluorescence lifetime is than 2~ns.}
\end{figure}

For the lifetime measurements, a Hamamatsu photomultiplier (R5509-43) was connected directly to the output of an optical fiber normally leading to the Raman spectrometer. The time resolution of the photomultiplier is about 2~ns, when using a 50~$\Omega$ input impedance. The laser signal was monitored with a Si photodiode. No significant difference between the time profile of the  laser pulse and the LIF signal was observed, Fig.~\ref{LIFPMT}. Therefore the lifetime of the electronically excited state is less than 2~ns. Such a short lifetime is  expected for the $^2A_{1u}$ state. from the linewidth observed for the allowed $^2A_{1u}\leftarrow ^2E_{1g}$ electronic transition of C$_{60}^+$, the gas-phase lifetime of the $^2A_{1u}$ state was estimated to be about 2~ps \cite{MaierDIBs2015}. The matrix environment probably makes it even shorter. A similar lifetime (2.2$\pm$0.2~ps) was found upon population of a vibronic state of the $^2E_g$ electronically excited state of C$_{60}^-$ by the fs-pump/probe photoelectron spectroscopy upon excitation at 775~nm \cite{C60anifs}. We therefore also explored whether the first excited state of matrix isolated C$_{60}^-$ (absorption maximum at 1066~nm in gas phase \cite{C60aniNIRGas}) shows measurable fluorescence. In matrixes also containing fullerene anions we did not observe a LIF signal from C$_{60}^-$ upon excitation of its near-IR vibronic bands.
\subsection*{Temperature dependence}
\begin{figure}
\includegraphics[width=0.4\textwidth]{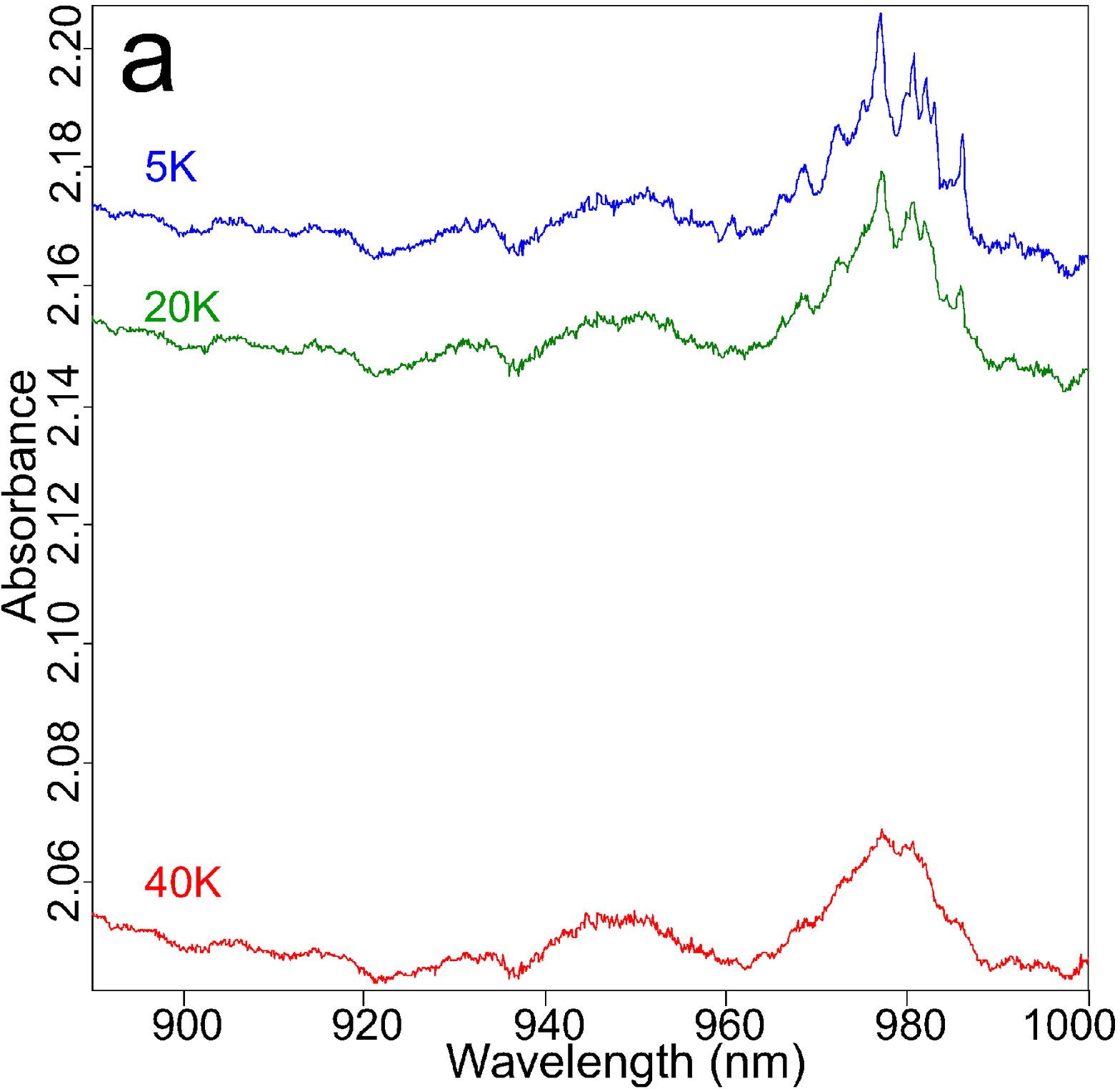}
\includegraphics[width=0.42\textwidth]{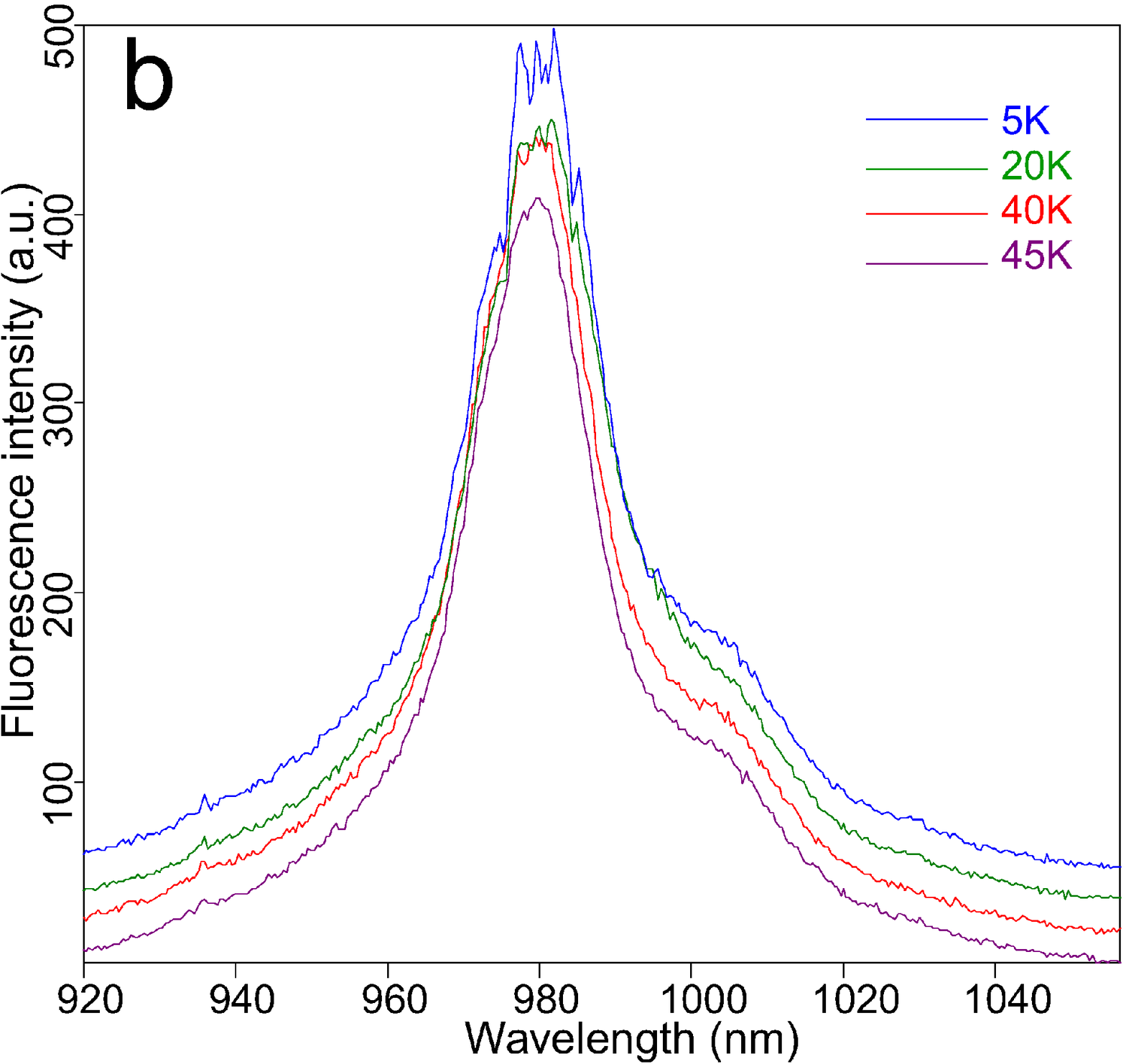}
\caption{\label{Kr} \textbf{(a)} Absorption spectra of C$_{60}^{+/-/0}$ in Kr matrix at various temperatures. The offset is real and is caused by changes of the optical properties of solid krypton, which lead to increased light scattering as the sample is cooled below 40~K. \textbf{(b)} LIF spectra of C$_{60}^{+/-/0}$ in Kr matrix at various temperatures (785~nm excitation). Original spectra have a similar baseline. A vertical offset was added to better distinguish the individual spectra.}
\end{figure}

Neon is not a convinient matrix within which to probe the temperature dependence of fluorescence processes, since Ne sublimes already at about 12~K. For such measurements we prepared a Kr matrix, containing C$_{60}^+$. The absorption spectrum of C$_{60}^+$ in Kr differs somewhat from that in Ne, see Fig.~\ref{Kr}a, but one can still measure pronounced LIF of C$_{60}^+$, Fig.~\ref{Kr}b. Increasing the matrix temperature influences the  optical properties of the solid. At higher temperatures solid krypton scatters less than at lower temperatures. This can be seen by the scattering offset of the absorption spectra which is induced by changing temperature (Fig.~\ref{Kr}a). Nevertheless, one does not observe a significant  change of the LIF signal. For temperatures up to 45~K, the integrated emission intensity decreases by less than 7\% (785~nm excitation). The spectra presented were obtained by depositing 8500~nAmin of C$_{60}^+$ at 5~K. Absorption and LIF measurements were done after annealing the matrix to 50~K (and subsequent cooling) to avoid possible changes of spectra due to diffusion of impurities or structural rearrangements of solid Kr. 

\begin{acknowledgments}
This work was supported by the Deutsche Forschungsgemeinschaft (KA 972/10-1). We also acknowledge support by KIT and Land Baden-W\"{u}rttemberg. 
\end{acknowledgments}

\bibliography{depo2}

\end{document}